\documentclass{optica-article}

\journal{opticajournal} 

\articletype{Research Article}

\usepackage{lineno}

\usepackage{graphicx}
\usepackage{bm}
\usepackage{siunitx}
\usepackage[caption=false]{subfig}
\usepackage{hyperref}
\usepackage{xcolor}
\usepackage{braket}
\usepackage{caption}
\usepackage{amsmath}

\begin{document}

\title{Entanglement Swapping with Integrated Narrowband Photon Sources for Quantum Repeaters}

\author{M.~Wu,\authormark{1,2} L.~Esmaeilifar,\authormark{3} R.~N.~Wang,\authormark{4} T.~J.~Kippenberg,\authormark{4} R.~Thew,\authormark{1,2} T. Brydges,\authormark{1,2,*}}

\address{\authormark{1}Department of Applied Physics, University of Geneva, Geneva, Switzerland\\
\authormark{2}Geneva Quantum Centre, University of Geneva, Geneva, Switzerland\\
\authormark{3}Department of Physics and Astronomy, Institute for Quantum Science and Technology, University of Calgary, AB T2N 1N4, Canada\\
\authormark{4}\'Ecole Polytechnique F\'ed\'erale de Lausanne (EPFL), Lausanne, Switzerland}

\email{\authormark{*}tiffany.brydges@unige.ch}

\begin{abstract*}
Promising implementations of first generation quantum repeaters are predicted to require atomic-based quantum memory systems interfaced with photonic sources. Integrated photonics provides a promising solution for fibre-based, field-deployed operation of quantum repeaters, however many leading quantum memory platforms require narrow-bandwidth photons that are challenging to generate with integrated photonics. Narrowband photons also present significant technical challenges when implementing entanglement-swapping, particularly with regards to systems-level stabilisation. This work addresses some of these fundamental and technical challenges, by demonstrating entanglement-swapping using state-of-the-art integrated photon sources with bandwidths compatible with multiple atomic-based quantum memory platforms. We obtained a background-subtracted (net) HOM visibility of 0.99$\,\pm\,$0.01, showing high photon indistinguishability and purity, with a net swapped state visibility of $\mathcal{V}=0.88\,\pm\,0.06$ demonstrating that the final entanglement would be sufficient to violate a Bell inequality. The experiment used independent pump lasers for each photon pair source, with highly different frequencies to mimic entanglement swapping between different repeater nodes or platforms. Phase and frequency stabilisation spanning 1.6\,THz was achieved using all-fibre, commercially-available components. These results address important challenges in implementing field-deployed quantum repeaters, from the integrated photonic solutions for narrowband photon pairs, to systems-level stabilisation between independent quantum repeater nodes.
\end{abstract*}

\section{Introduction}

\begin{figure}[t]
\centering
\includegraphics[width=.99\linewidth]{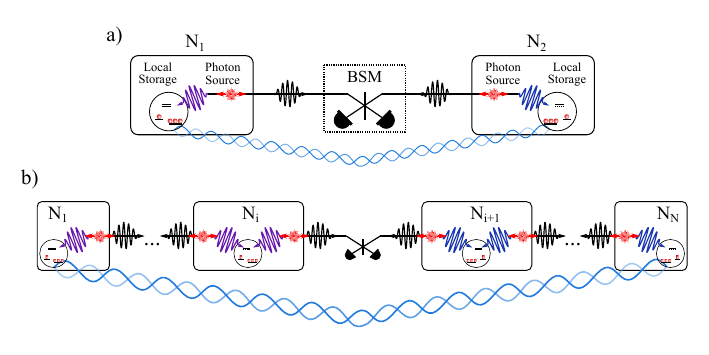}
\caption{Simplified schematic of one implementation of a quantum repeater network. a) An elementary quantum repeater link in the network. Nodes $N_{1}$ and $N_{2}$ prepare an entangled resource state. A part of this state is sent through a quantum channel to an intermediate Bell State Measurement (BSM) station, with the remaining part stored locally \cite{Briegel:1998, Duan:2001, Sangouard:2011}. Entanglement is distributed between $N_{1}$ and $N_{2}$ through a BSM, which projects the remaining states at $N_{1}$ and $N_{2}$ into an entangled state. b) Schematic of full repeater link, allowing nodes $N_{1}$ and $N_{N}$ to share an entangled state through the use of additional nodes.}
\label{fig:Overview}
\end{figure}

 Through the distribution of entanglement, quantum networks are envisaged to be crucial for enabling many different quantum technologies. Short distance networks which can connect (potentially disparate) subsystems are already relevant for distributed quantum computing \cite{Jiang:2007,Main:2025,Barral:2025}, with global quantum networks of significance for secure quantum communication \cite{Kimble:2008, Wehner:2018} and various quantum sensing applications \cite{Marchese:2023, Stas:2026}. However, the loss of typical fibre networks places practical distance limits on the full implementation of quantum networks \cite{Takeoka:2014, Pirandola:2017, Pirandola:2020, Liu:2023}. A solution to this is the quantum repeater, which distributes entanglement between users via repeater nodes \cite{Sangouard:2011, Azuma:2023}. A cornerstone of quantum repeater architectures is the entanglement swapping protocol, where entanglement is teleported between two parties by performing joint measurements on their entangled partners \cite{Zukowski:1993}. A schematic representation of an elementary quantum repeater link utilising entanglement-swapping can be seen in Fig.~\ref{fig:Overview} a). Two nodes, $N_{1}$ and $N_{2}$, each prepare an entangled resource state, part of which is sent through a quantum channel to an intermediate station. A joint ``Bell state measurement'' (BSM), performed at the intermediate station, projects the remaining states at $N_{1}$ and $N_{2}$ into an entangled state. For large-scale networks, both a resource of entanglement along with light-matter interfaces are anticipated to be necessary \cite{Briegel:1998, Duan:2001}.\\

First generation quantum repeaters require the use of quantum memories \cite{Sangouard:2011, Muralidharan:2016, Azuma:2023}, which allow the storage and retrieval of quantum states \cite{Tittel:2010}. Some of the most promising quantum memories are those based on absorptive solid-state atomic systems for single-photon storage and retrieval \cite{Tittel:2025}, however typically such memories have absorption profiles with a narrow bandwidth in comparison to many standard probabilistic photon sources. This usually requires modification of the source to include an external cavity to achieve the appropriate bandwidth, leading to highly complex setups requiring large amounts of bulk optics \cite{Rivera:2021, Hanni:2025, Sanchez:2026}, which is impractical when moving to field-deployed quantum networks. In allowing large numbers of components to be packaged in a compact manner, integrated photonics is a promising solution \cite{Moody:2022}, motivating the development of integrated photon sources with quantum memory compatible bandwidths.\\

Previously, integrated silicon nitride (SiN) micro-ring resonators (MRRs) have been shown to produce high quality energy-time entangled photon pairs for use as an entanglement resource in quantum networks \cite{Llewellyn:2020, Samara:2021, Chen:2024}. However, previous entanglement-swapping demonstrations used sources with bandwidths significantly broader than those compatible with atomic-based quantum memories. Proof-of-principle teleportation using a MRR and a solid-state quantum memory has been recently demonstrated \cite{An:2025}. However, this used a weak coherent state for demonstration of teleportation, along with significantly complex bulk optics for frequency stabilisation, and a photon source still too broadband for compatibility with many leading solid-state quantum memory platforms \cite{Sanchez:2026}.\\

The work presented here shows proof-of-concept entanglement swapping between two independent, state-of-the-art integrated photon-pair sources with narrow bandwidths, compatible with multiple atomic quantum memories based on solid-state systems \cite{An:2025, Sanchez:2026}. The experiment uses an integrated source and fibre-optics to demonstrate the suitability of the source and setup for future deployed quantum networks, and implements frequency and phase stabilisation over a span of \SI{1.6}{THz} using only simple, off-the-shelf components. The photon-pair sources generate energy-time entangled photon pairs, with bandwidths $<$\SI{60}{MHz} which corresponds to coherence times $>$\SI{5}{ns} (loaded Q-factors $>$$\num{3e6}$). A BSM at a central station and post-selection of events separated by time $\tau$ projects the remaining photons into a time-bin entangled state \cite{Halder:2007}. The long coherence time of the photons, which exceeds the temporal resolution of the detectors, allows the photon timing to be determined by the detection events, so selecting out a single temporal mode and ensuring high spectral purity \cite{Huang:2010}. However, this feature also introduces significant technical challenges in comparison to previous entanglement swapping schemes \cite{Samara:2021}, such as strict frequency stabilisation between the two nodes. This is, to the authors' knowledge, the first entanglement swapping demonstration with two integrated photon sources of $<$\SI{60}{MHz} bandwidth. The results address important technical challenges in systems-level stabilisation between independent repeater nodes, and are a key step towards implementing field-deployed quantum repeaters.

\begin{figure*}[t]
\centering
 \makebox[\textwidth][c]{\includegraphics[width=1.25\textwidth]{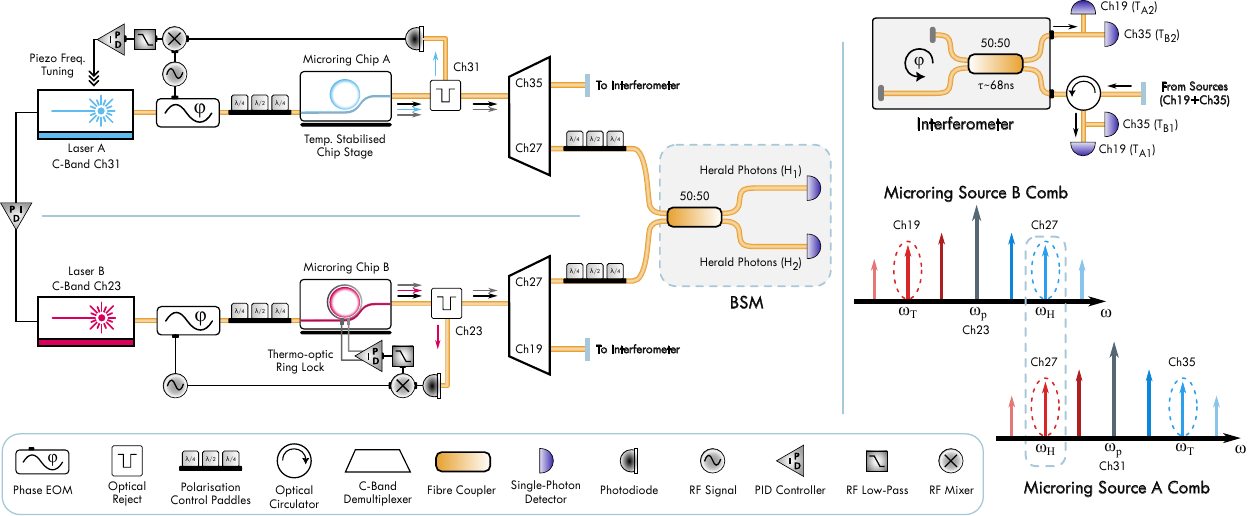}}
\caption{Simplified schematic of the experimental setup. The general cavity-lock feedback loops are shown for each source. The photon pairs from each source are demultiplexed according to the ITU C-band frequency grid: Ch27 photons are interfered at the central BSM station, with the remaining Ch19 and Ch35 photons passed through a folded Franson interferometer for the time-bin entanglement measurement. A detailed schematic of the experimental setup is given in the supplementary material.}
\label{fig:EntSwappoverview}
\end{figure*}

\section{Experimental Setup and Implementation}

A schematic overview of the main experimental setup is shown in Fig~\ref{fig:EntSwappoverview}, with a more detailed schematic and further information given in the Supplementary Material.

\subsection{Entangled Photon Sources}

Each photon source, designated A and B, was comprised of a  state-of-the-art high-Q MRR pumped by a continuous-wave (CW) laser source. The MRRs are SiN waveguides with a $\text{SiO}_{2}$ cladding, fabricated using the photonic Damascene process \cite{Pfeiffer:2016}. Both MRRs were designed with a free-spectral range (FSR) of $\sim$\SI{200}{\giga\hertz} to be compatible with the ITU C-band frequency grid. Additionally, MRR-B contained heater electrodes for direct tuning of the on-chip cavity resonance frequency through an applied current via the thermo-optic effect, enabling MRR-B to be actively frequency-stabilised to laser B through the Pound-Drever-Hall (PDH) technique \cite{Drever:1983}. Conversely, laser A was actively stabilised to MRR-A, also through a PDH lock. All feedback systems in this work were implemented using low-noise commercial RF System-on-Chip (RF-SoC) devices (Red Pitaya Stemlab125-14) combined with home-built amplifier boards. Lasers A and B have a frequency difference of $\sim$\SI{800}{\giga\hertz}, chosen to mimic future quantum network implementations where independent nodes or platforms may operate at very different frequencies. Energy-time entangled photon-pairs were generated from each source through cavity-enhanced spontaneous four-wave mixing (SFWM) \cite{Samara:2019} with the resulting frequency comb aligned to the C-band frequency channels. Table \ref{tab:source_freq} lists the central wavelengths and frequency linewidths of each mode, where we designate \textit{herald} as the photons which interfere at the central beamsplitter for a BSM, and \textit{target} as the heralded photons onto which entanglement is swapped.

\begin{table}[htbp]
\centering
\caption{\bf Characteristics of each optical mode from the photon-pair sources}
\label{tab:source_freq}
\begin{tabular}{cccc}
\hline
\bf Mode & \bf Wavelength (\SI{}{\nano\meter})& \bf C-Band Channel & \bf Linewidth (\SI{}{\mega\hertz}) \\ 
\hline
Source A Herald  & 1555.75 & Ch27 & 51 \\
Source A Target  & 1549.32 & Ch35 & 56 \\
Source B Herald  & 1555.75 & Ch27 & 39 \\
Source B Target  & 1562.23 & Ch19 & 37 \\

\hline
\end{tabular}
\end{table}

\subsection{Entanglement Swapping}
In the CW regime, highly time-correlated photon-pairs are generated at indeterminate times within the coherence of the pump (which is much larger than that of the photons). Consequently, the photon-pair state for source A(B) can be represented as a bipartite coherent superposition of all possible generation times such that $|\psi\rangle_{A(B)} \propto \sum_{t} |t,t\rangle_{A_HA_T(B_HB_T)}$. The state of the two-source system is then the tensor product $|\psi\rangle_{A}\otimes|\psi\rangle_{B}$, and for final analysis in the time-bin basis it is instructive to represent this state as a discretised superposition of generation times:

\begin{align}
\begin{split}
\ket{\psi} = 
\ket{\psi}_{A}\otimes\ket{\psi}_{B}\propto 
\sum_{t}&\biggl(\ket{t,t}_{A_HA_T}\ket{t,t}_{B_HB_T} +
\\ &\sum_{\tau>0}(\ket{t,t}_{A_HA_T}\ket{t+\tau,t+\tau}_{B_HB_T}+\ket{t+\tau,t+\tau}_{A_HA_T}\ket{t,t}_{B_HB_T})\biggr).
\end{split}
\end{align}
The first term in the summation describes a pair of photons being created simultaneously at each source (i.e. four simultaneous photons), with the latter term describing a photon pair created at each source separated by a time delay of $\tau$, with $\tau_c\gg\tau>0$ where $\tau_c$ is the coherence time of the laser pumps. The photons at Ch27 from each pair source (heralds) were interfered at a central BSM station comprised of a fibre 50/50 beam-splitter, with the outputs subsequently detected by two Superconducting Nanowire Single Photon Detectors (SNSPDs), $H_1$ and $H_2$ \cite{Autebert:2020}, as shown in Fig~\ref{fig:EntSwappoverview}. Detector clicks at time $t_0=t$ and $t_1=t+\tau$ for some $\tau$ at $H_1$ and $H_2$ are equivalent to a measurement in the Bell state $\ket{\Psi^{-}}\propto(\ket{t_0,t_1}_{HH}-\ket{t_1,t_0}_{HH})$ \cite{Brendel:1999}. A successful click pattern ``heralds" entanglement swapping, with the remaining target photons from each source subsequently projected onto the time-bin entangled state $\ket{\Psi^{-}}_{TT}\propto(\ket{t_0,t_1}_{TT}-\ket{t_1,t_0}_{TT})$, even though they have never interacted. As the photons were created at arbitrary times, post-selection ensures only those photons which were separated in time by $\tau$ are considered successful events, thus removing the need for timing-synchronisation between the pump lasers as with a pulsed configuration.

The detection times at the two time-bins were well-defined relative to the coherence time of the photons, due to the low jitter of the SNSPDs ($\sim$\SI{50}{ps}). For simplicity, in this work a partial BSM was considered where only the state $|\Psi^{-}\rangle$ was heralded. Due to working in the energy-time regime, it was necessary to ensure only a single temporal mode was heralded at the BSM, through choosing a coincidence window sufficiently narrow compared to the coherence time of the photons \cite{Huang:2010,Baghdasaryan:2026}. Further discussion of the impact of coincidence windows on the visibility is given in the Supplementary Material.

\subsection{Verification of the swapped state}
For analysis of the swapped state, the target photons Ch35 and Ch19 were recombined into the same path and sent to a single unbalanced interferometer with time-delay $68.7\,\pm\,$\SI{0.1}{\nano\second}. This is sufficiently longer than the coherence time of the single photons ($\sim$\SI{5}{ns}), in order to avoid single-photon interference which will degrade the two-photon interference visibility measurement, and also leaves reasonable margin for more narrowband photons in future experiments. Depending on the click patterns at the output of the interferometer, either a Z-basis measurement or a superposition-basis measurement was performed \cite{Singh:2025}. A common interferometer was used for this as a proof-of-principle demonstration of entanglement swapping with narrowband photons. As the two target photons were of very different wavelengths, each photon essentially saw its own individual interferometer. Additionally, as shown in Fig.~\ref{fig:Overview} a), future first generation quantum networks will use quantum memories at each node for both storage and interferometric readout \cite{Clausen:2011}. For the absorptive solid state quantum memories this work considers, such memories offer high interferometric stability in comparison to fibre Franson interferometers \cite{Kutluer:2019}, with a single Franson interferometer being a better representation of this highly stable scheme than two separate interferometers.

The interferometer was actively phase-stabilised to laser A using a locking scheme with an external EOM \cite{Rogers:2016}. For entanglement swapping using two independent sources with independent pump lasers, a measurement of the heralding photons in the time-bin state $|\psi\rangle_{A_HB_H} \propto |t_{0},t_{1}\rangle_{A_HB_H} - |t_{1},t_{0}\rangle_{A_HB_H}$ projects the remaining photons into the state

\begin{align}\label{eq:heralded_swap}
\ket{\psi}_{A_TB_T}\propto\ket{t_{0},t_{1}}_{A_TB_T} - \mathrm{e}^{i(\Delta\phi_{A}-\Delta\phi_{B})}\ket{t_{1},t_{0}}_{A_TB_T},
\end{align}
where $\Delta\phi_{A(B)}$ is the phase difference between the two time-bins of Source A(B). Passing the target photons through the folded Franson interferometer with the same $\tau$ imbalance, and selecting only those events with coincident detections, transforms the state to
\begin{align}\label{eq:_heralded_coin_swap}
\ket{\psi}'_{A_TB_T} \propto \left(1-\mathrm{e}^{i(\Delta\phi_A - \Delta\phi_B-\phi_\alpha - \phi_\beta)}\right)\ket{t_1, t_1}_{A_TB_T},
\end{align}
where $\phi_{\alpha(\beta)}$ is the phase shift/setting applied by the Franson interferometer to the Source A(B) photon in the target streams (derivation and further discussion are given in the Supplementary Material). As such, changes in the frequency between the pump lasers directly change (for a fixed $\tau$) the phase of the entangled state onto which the heralded photons are projected. Hence, the frequency difference between lasers A and B must be stabilised during a measurement. This was achieved by passing a portion of the light from laser B into the interferometer and locking laser B's frequency to the resulting phase error signal: The interferometer doubled as a frequency/phase ``bridge'', since the interferometer was in turn phase-locked to laser A. 

In particular, the resulting frequency difference between the photons of the swapped state was $\sim$\SI{1.6}{THz}, thus through using the interferometer we were able to achieve frequency/phase stability across a large frequency difference without the need for bulky, high-finesse cavities, which have been used in previous teleportation implementations \cite{An:2025}. For future quantum network implementations with spatially separated nodes, a single SI-traceable signal distributed to all nodes in the network \cite{Husmann:2021} can be used, to which each node can stabilise all laser systems to ensure a common frequency is shared over the network.

Having controlled for the phase of the swapped state, we measured coincidences between the target photons from the output ports of the interferometer. This is a time-bin measurement in the two-qubit superposition basis, with the output port permutations corresponding to different joint measurements in that basis. With reference to Figure \ref{fig:EntSwappoverview}, the probabilities for coincidence clicks between the different modes can be shown to be: $\Pr(T_{A1}T_{B1}) = \Pr(T_{A2}T_{B2}) \propto 1-\cos(\Delta\phi)$ and $\Pr(T_{A2}T_{B1}) =\Pr(T_{A1}T_{B2}) \propto 1+\cos(\Delta\phi) \equiv 1-\cos(\Delta\phi+\pi)$ for some total phase offset $\Delta\phi$. These two outcomes correspond to measurements of the visibility curve which are $\pi$-out of phase with one-another. As such, when measuring the swapped state visibility, two phase positions are measured simultaneously for a single basis choice. The phase setting itself was scanned via changing the driving frequency of an acousto-optic modulator (AOM) placed in the path of the locking laser light from laser B before the interferometer, which had the effect of shifting laser B's frequency relative to laser A. Full details of the coincidence measurement and the AOM-based phase control can be found in the Supplementary Material.

\begin{figure}[t]
\centering
\includegraphics[width=.99\linewidth]{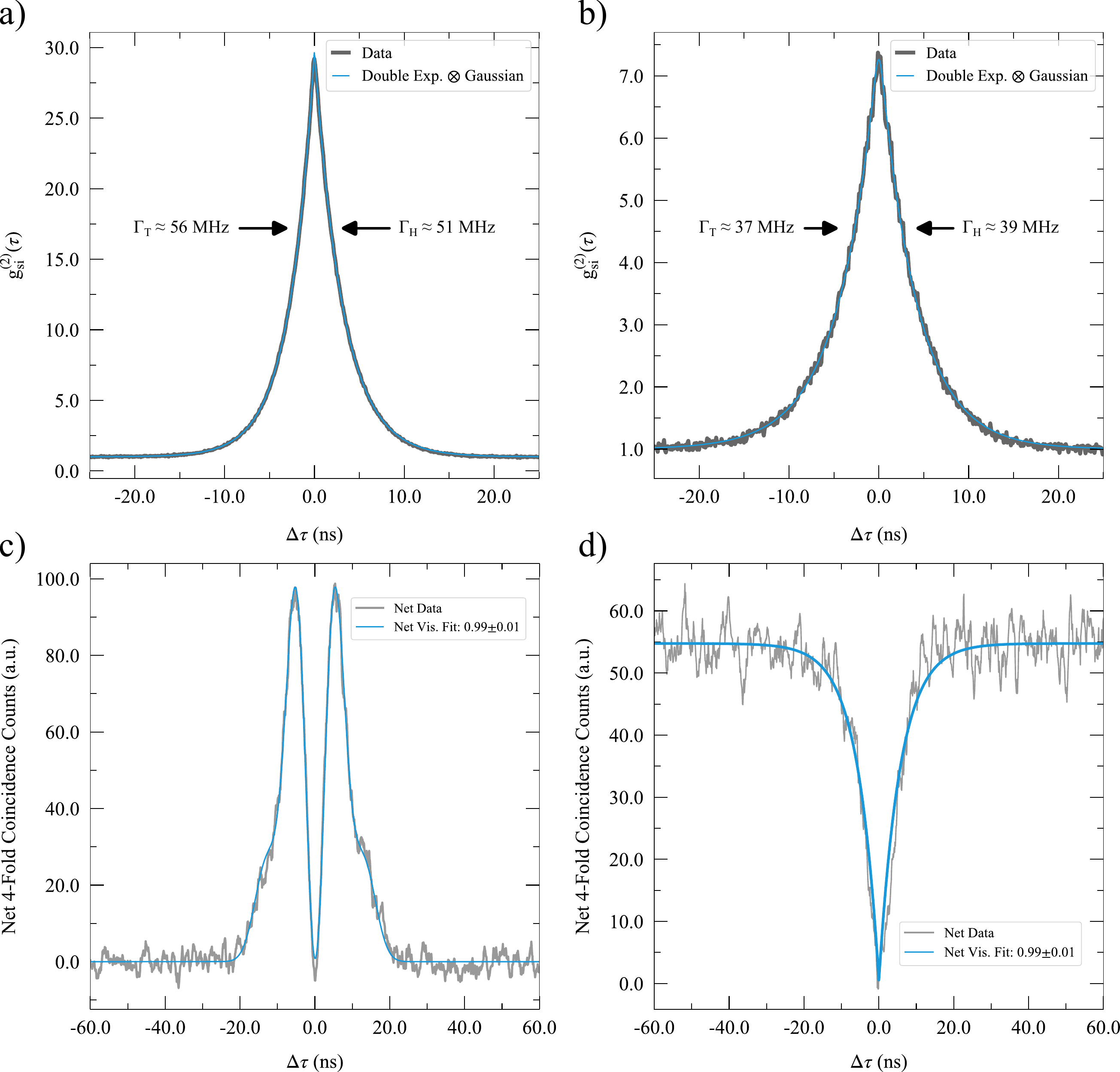}
\caption{a) and b) Cross-correlation measurements for Source A and Source B, respectively. Data is shown in black, with the blue lines fits to the data from a convolution of a double exponential with a Gaussian. The extracted photon linewidths are: $\Gamma_{T_A}\approx56\,$MHz and $\Gamma_{H_A}\approx51\,$MHz (Source A), $\Gamma_{T_B}\approx37\,$MHz and $\Gamma_{H_B}\approx39\,$MHz (Source B). c) and d) Background-subtracted (net) HOM interference measurements with coincidence window of $\pm$\SI{15}{ns} and $\pm$\SI{0.7}{ns} respectively. Both measurements give a fitted net visibility of 0.99\,$\pm$\,0.01. For all four plots, the data is given as thick gray lines and the fits are thin blue lines (see Supplementary Material for the fit models).}
\label{fig:HOMs}
\end{figure}

\section{Results}
\subsection{Photon Bandwidths}
Cross-correlation measurements were taken to directly determine the bandwidths of all four photons, with the results shown in Fig.~\ref{fig:HOMs} a) and b) for Source A and Source B respectively. A fit to the data, comprising a double exponential convolved with the Gaussian jitter of the detectors, gives bandwidths of: $\Gamma_{T_A}\approx56\,$MHz and $\Gamma_{H_A}\approx51\,$MHz (Source A), $\Gamma_{T_B}\approx37\,$MHz and $\Gamma_{H_B}\approx39\,$MHz (Source B).

\subsection{Photon Indistinguishability}
For a high-fidelity BSM, the photons arriving at the beam-splitter must be indistinguishable in all degrees of freedom \cite{tsujimoto2017, Baghdasaryan:2026}. The most common method of determining the indistinguishability of independent photons is through Hong-Ou-Mandel (HOM) interference \cite{Hong:1987}. In particular, when the coherence time of the photons is much larger than the detector resolution, HOM interference can take place in two different regimes, depending on whether frequency information can be resolved. In the frequency-resolved regime, the coincidence window, $\Delta\tau$, is on the same order as the coherence time of the photons, and so -- coupled with the detector jitter being much smaller than the coherence time of the photons -- frequency differences between the two interfering photons can be resolved. This is seen as a quantum beating feature in the heralded HOM interference which is proportional to the frequency detuning between the two photons \cite{Legero:2003, Legero:2004, An:2025}. Fig.~\ref{fig:HOMs} c) shows this frequency-resolved HOM interference, where the classic HOM dip is present at $\Delta\tau=0$, however anti-bunching occurs as the time difference between detections increases (with the background extracted from the fit and subsequently subtracted). The decay from this anti-bunching peak occurs due to the cut-off from the coincidence window, here chosen to be $\pm\SI{15}{ns}$. An increase in the detuning leads to a faster beating structure and vice versa. Since this effect was also present with the \textit{unheralded} HOM measurement, we were able to use fast ($\sim$\SI{10}{\second}) 2-fold coincidence measurements to align the herald photons' central frequencies to within $\sim$\SI{5}{\mega\hertz} (see Supplementary Material). The fitted net visibility of the quantum beat is 0.99\,$\pm$\,0.01 demonstrating high indistinguishability between the photons in all degrees of freedom except frequency (resolved by the detection). Conversely, by reducing the coincidence window to be much smaller than the photon coherence time, thus temporally filtering the photons, the frequency uncertainty of the detection becomes much larger than the photon bandwidths. In this non-frequency-resolved regime, the HOM interference returns to the more commonly known shape as in Fig.~\ref{fig:HOMs} d), where no beating (and no anti-bunching) is resolvable. Fig.~\ref{fig:HOMs} d) shows the background-subtracted HOM interference for a coincidence window of $\pm\SI{0.7}{ns}$, where a fit to the data retrieves a similar net visibility of 0.99\,$\pm$\,0.01.

Background subtraction was necessary with these devices due to a significant level of background counts caused predominantly by noise photons from Raman scattering in the $\text{SiO}_{2}$ cladding of both MRR-A and MRR-B \cite{Samara:2019, Pal:2025}. Raman scattering is particularly enhanced in these devices compared to previous work due to the very high Q-factor of the resonators. This causes the intra-cavity power to be extremely high compared to resonators with lower Q-factors, leading to cavity-enhanced Raman scattering effects in the cavity itself. For further information on the background, including the methodology for background subtraction, see the Supplementary Material. This background could be reduced by exploring the use of alternative claddings, such as air \cite{Moille:2025}. However, the outlook of this work is to generate narrowband \textit{and} highly non-degenerate photon pairs from integrated MRRs for interfacing with atomic quantum memory systems in the visible/NIR regime, where the frequency overlap between the Raman scattering (near the pump) and the photon pairs will be minimal.

\subsection{Verification of Entanglement Swapping}

\begin{figure}[t]
\centering
\includegraphics[width=1.\linewidth]{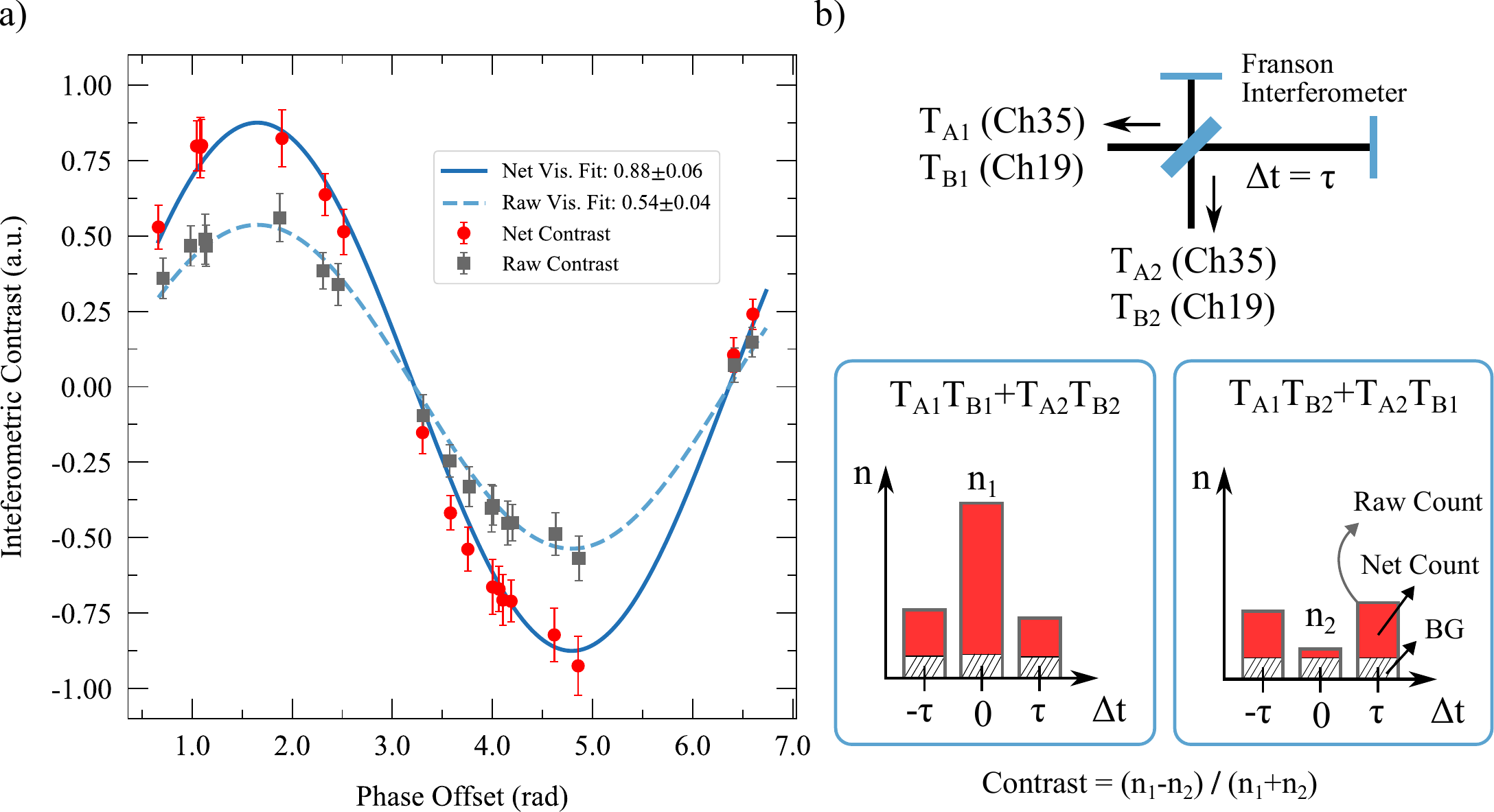}
\caption{a) Interferometric contrast of the entanglement swapping measurement as a function of the phase setting of the swapped state, with the BSM and target photon coincidence windows being $\pm\SI{0.7}{\nano\second}$ and $\pm\SI{1.5}{\nano\second}$ respectively. The raw visibility is 0.54\,$\pm$\,0.04, with a background subtracted visibility of 0.88\,$\pm$\,0.06, with the reduced chi-squared for the net fit being $\chi^2_\nu=1.58$. b) Conceptual diagrams of how coincidences are measured between the two sources' target photon streams and used to calculate the interferometric contrast for each phase setting. For the C-band Ch35 (Source A) and Ch19 (Source B) photon modes, they each have two interferometer output signal streams, giving four permutations of time-bin coincidence measurements representing two maximally anti-correlated coincidence counts $n_1$ and $n_2$, which we can use to calculate the contrast as $(n_1-n_2)/(n_1+n_2)$.}
\label{fig:SwappedVis}
\end{figure}

During a single measurement when characterising the swapped state, fluctuations in the photon rates occurred due to power fluctuations predominantly from drifts in the coupling to the integrated chips. Directly plotting the photon rates as a function of interferometer phase would therefore lead to a visibility curve modulated by power fluctuations, rather than being a direct measurement of the swapped state. To construct a parameter independent of power fluctuations, we can exploit the fact that we simultaneously measure the two maximally anti-correlated phase positions for each phase offset. Fig.~\ref{fig:SwappedVis} b) shows an example: $n_{1}$ was extracted from coincidences between modes $T_{A1}T_{B1}$ summed with that of $T_{A2}T_{B2}$ (common-port events), with $n_{2}$, the $\pi$-shifted measurement, being the sum of $T_{A1}T_{B2}$ and $T_{A2}T_{B1}$ (cross-port events). From $n_{1}$ and $n_{2}$, we can construct the two-photon normalised correlation coefficient, which we also term the interferometric contrast $\mathcal{I}$, where $\mathcal{I} = (n_{1}-n_{2})/(n_{1}+n_{2})$ \cite{aspect82, Halder:2007}. The contrast is bounded between -1 and 1, and is insensitive to first order impacts on the visibility from power fluctuations. We measured the contrast while scanning the aggregate phase of the swapped state as shown in Fig.~\ref{fig:SwappedVis} a) and fitted the data via

\begin{equation}
    \mathcal{I}(\Delta\phi) = \mathcal{V}\cos(\Delta\phi+\phi_{0}),
    \label{eq:visibility_fit}
\end{equation}
where $\Delta\phi$ is the scanned phase of the swapped state, $\mathcal{V}$ is the fitted visibility parameter, and $\phi_0$ is a fixed phase offset. We note that due to long integration times ($\sim$\SI{3}{\hour} per phase setting), the data points were taken in smaller groups across several days, with each group having a different absolute phase position since the interferometer is relocked each session. As such, we fitted Eq.~\ref{eq:visibility_fit} as a piecewise function with multiple $\phi_0$ values corresponding to the measurement sessions (details in Supplementary Material). For BSM and time-bin analysis coincidence windows of $\pm\SI{0.7}{\nano\second}$ and $\pm\SI{1.5}{\nano\second}$ respectively, we observed an average 4-fold coincidence rate of $\sim$0.03\,\SI{}{\hertz}. The non-background-subtracted (raw) visibility was found to be $\mathcal{V}_{raw}=0.54\,\pm$\,0.04, with the background-subtracted (net) visibility being $\mathcal{V}_{net}=0.88\,\pm$\,0.06. Both $\mathcal{V}_{raw}$ and $\mathcal{V}_{net}$ exceed $1/3$, indicating the swapped photons are in a non-separable state \cite{Peres:1996}. As $\mathcal{V}_{net}$ additionally exceeds $1/\sqrt{2}$, the swapped entanglement is sufficient to violate a Bell inequality \cite{Clauser:1969}. Further discussion on time-bin entanglement and the background-subtraction procedure is found in the Supplementary Material.

We may estimate the expected swapped state visibility by modelling it as the direct product of the constituent visibilities

\begin{equation}
\mathcal{V}_{swap} = \mathcal{V}_{ET_{1}}\cdot \mathcal{V}_{ET_{2}}\cdot\mathcal{V}_{HOM}\cdot\mathcal{V}_\phi,
\end{equation}
where $\mathcal{V}_{ET_{1,2}}$ is the energy-time entanglement visibility of each source, and $\mathcal{V}_{\phi}$ is the phase stability between the two laser sources which depends on the interferometer frequency bridge stability. The average visibility of the swapped state is upper-bounded by the product of the visibilities of the individual sources \cite{zangi2023, zukowski2005}, while $\mathcal{V}_{HOM}$ and $V_{\phi}$, being measures of the independent fidelities of the BSM and swapped state time-bin measurement respectively, additionally bound the maximum measurable $\mathcal{V}_{swap}$ \cite{tsujimoto2017}. With background subtraction, $\mathcal{V}_{ET_1}$ ($\mathcal{V}_{ET_2}$) was measured to be $0.954\pm0.001$ ($0.939\pm0.001$); we attribute the non-unity visibilities mainly to imperfections arising from the highly imbalanced interferometer. Finally, by analysing the phase lock error signal over the duration of the swapping measurement, we estimated $\mathcal{V}_\phi\sim0.994$ (see Supplementary Material), giving a net expected value of $\mathcal{V}_{swap}\sim0.882$, in good agreement with the measured result.

\section{Summary and Discussion}

This work is the first implementation of entanglement swapping using narrowband photons generated by integrated sources, with bandwidths compatible with quantum memories based on solid-state systems, and using predominantly fibre-based, off-the-shelf components -- an important step forward for the development of deployed quantum networks. The state-of-the-art integrated MRR sources generated photons with bandwidths $<$60\,MHz, compatible with multiple solid-state quantum memories. This work addressed many of the technical challenges which arise from using such narrowband photons, without the use of bulky, complex components such as high-finesse cavities which have been used in previous demonstrations. A flexible architecture was implemented which used independent pump lasers of very different wavelengths to mimic different repeater nodes and platforms, requiring phase and frequency stabilisation over \SI{1.6}{THz}. Despite this, high photon indistinguishability was shown through a net HOM visibility of $0.99\pm0.01$, with measured net and raw entanglement swapping visibilities of $0.88\pm0.06$ and $0.54\pm0.04$ respectively, clearly demonstrating non-separable states and demonstrating that the final entanglement would be sufficient to violate a Bell inequality. This work is an important step towards enabling the development of future deployed quantum networks using integrated photon sources interfaced with quantum memories based on solid-state systems. It highlights some of the important challenges, including stabilisation of the different nodes to a shared, stable distributed frequency reference. Future work will move towards interfacing narrowband, highly non-degenerate photon pairs with atomic quantum memory systems in the visible/NIR \cite{Zektzer:2024}, combined with interferometric readout to enable verification of entanglement swapping.

\begin{backmatter}
\bmsection{Funding}
T.B. and M.W. are supported by the Swiss National Science Foundation through Ambizione Grant No. PZ00P2$\_$216153. This work was supported by the Swiss State Secretariat for Education, Research and Innovation (SERI) (Contract No. UeM019-3). L.E. was supported by the Mitacs Globalink Research Award.

\bmsection{Acknowledgment}
The authors thank David Cabrerizo for assistance with the in-house electronics development, and Nikolai Kuznetsov for providing the micro-ring resonators. We also thank F. Joseph Marcellino for useful discussions. 

\bmsection{Disclosures}
The authors declare no conflicts of interest.

\bmsection{Data availability} Data underlying the results presented in this paper are not publicly available at this time but may be obtained from the authors upon reasonable request.

\bmsection{Supplemental document}
See Supplement 1 for supporting content.

\end{backmatter}

\appendix


\bibliography{Bibliography}

@article{An:2025,
  title = {Quantum Teleportation from Telecom Photons to Erbium-Ion Ensembles},
  author = {An, Yu-Yang and He, Qian and Xue, Wenyi and Jiang, Ming-Hao and Yang, Chengdong and Lu, Yan-Qing and Zhu, Shining and Ma, Xiao-Song},
  journal = {Phys. Rev. Lett.},
  volume = {135},
  issue = {1},
  pages = {010804},
  numpages = {8},
  year = {2025},
  month = {Jul},
  publisher = {American Physical Society},
  doi = {10.1103/3wh8-2gh1},
  url = {https://link.aps.org/doi/10.1103/3wh8-2gh1}
}

@article{aspect82,
  title = {Experimental Realization of Einstein-Podolsky-Rosen-Bohm Gedankenexperiment: A New Violation of Bell's Inequalities},
  author = {Aspect, Alain and Grangier, Philippe and Roger, G\'erard},
  journal = {Phys. Rev. Lett.},
  volume = {49},
  issue = {2},
  pages = {91--94},
  numpages = {0},
  year = {1982},
  month = {Jul},
  publisher = {American Physical Society},
  doi = {10.1103/PhysRevLett.49.91},
  url = {https://link.aps.org/doi/10.1103/PhysRevLett.49.91}
}

@article{Autebert:2020,
author = {C. Autebert and G. Gras and E. Amri and M. Perrenoud and M. Caloz and H. Zbinden and F. Bussi\`res},
journal = {J. Appl. Phys.},
pages = {074504},
publisher = {AIP Publishing},
title = {{Direct measurement of the recovery time of superconducting nanowire single-photon detectors}},
volume = {128},
year = {2020},
url = {https://doi.org/10.1063/5.0007976},
doi = {10.1063/5.0007976},
}

@article{Azuma:2023,
  title = {Quantum repeaters: From quantum networks to the quantum internet},
  author = {Azuma, Koji and Economou, Sophia E. and Elkouss, David and Hilaire, Paul and Jiang, Liang and Lo, Hoi-Kwong and Tzitrin, Ilan},
  journal = {Rev. Mod. Phys.},
  volume = {95},
  issue = {4},
  pages = {045006},
  numpages = {66},
  year = {2023},
  month = {Dec},
  publisher = {American Physical Society},
  doi = {10.1103/RevModPhys.95.045006},
  url = {https://link.aps.org/doi/10.1103/RevModPhys.95.045006}
}

@article{tsujimoto2017,
author = {Yoshiaki Tsujimoto and Yukihiro Sugiura and Motoki Tanaka and Rikizo Ikuta and Shigehito Miki and Taro Yamashita and Hirotaka Terai and Mikio Fujiwara and Takashi Yamamoto and Masato Koashi and Masahide Sasaki and Nobuyuki Imoto},
journal = {Opt. Express},
number = {11},
pages = {12069--12080},
publisher = {Optica Publishing Group},
title = {High visibility Hong-Ou-Mandel interference via a time-resolved coincidence measurement},
volume = {25},
month = {May},
year = {2017},
url = {https://opg.optica.org/oe/abstract.cfm?URI=oe-25-11-12069},
doi = {10.1364/OE.25.012069},
}

@article{zukowski2005,
  title = {Entanglement swapping of noisy states: A kind of superadditivity in nonclassicality},
  author = {Sen(De), Aditi and Sen, Ujjwal and Brukner, \ifmmode \check{C}\else \v{C}\fi{}aslav and Bu\ifmmode \check{z}\else \v{z}\fi{}ek, Vladim\'{\i}r and \ifmmode \dot{Z}\else \.{Z}\fi{}ukowski, Marek},
  journal = {Phys. Rev. A},
  volume = {72},
  issue = {4},
  pages = {042310},
  numpages = {10},
  year = {2005},
  month = {Oct},
  publisher = {American Physical Society},
  doi = {10.1103/PhysRevA.72.042310},
  url = {https://link.aps.org/doi/10.1103/PhysRevA.72.042310}
}

@Article{zangi2023,
AUTHOR = {Zangi, Sultan M. and Shukla, Chitra and ur Rahman, Atta and Zheng, Bo},
TITLE = {Entanglement Swapping and Swapped Entanglement},
JOURNAL = {Entropy},
VOLUME = {25},
YEAR = {2023},
NUMBER = {3},
ARTICLE-NUMBER = {415},
URL = {https://www.mdpi.com/1099-4300/25/3/415},
PubMedID = {36981304},
ISSN = {1099-4300},
DOI = {10.3390/e25030415}
}

@article{Baghdasaryan:2026,
author = {B. Baghdasaryan and K. Lozano-M\'endez and M. Leipe and M. Cabrejo-Ponce and S. H\{a}ussler and K. Joarder and T. G\{u}

@article{Barral:2025,
title = {{Review of Distributed Quantum Computing: From single QPU to High Performance Quantum Computing}},
journal = {Computer Science Review},
volume = {57},
pages = {100747},
year = {2025},
issn = {1574-0137},
doi = {https://doi.org/10.1016/j.cosrev.2025.100747},
url = {https://www.sciencedirect.com/science/article/pii/S1574013725000231},
author = {David Barral and F. Javier Cardama and Guillermo Díaz-Camacho and Daniel Faílde and Iago F. Llovo and Mariamo Mussa-Juane and Jorge Vázquez-Pérez and Juan Villasuso and César Piñeiro and Natalia Costas and Juan C. Pichel and Tomás F. Pena and Andrés Gómez}
}

@article{Briegel:1998,
  title = {Quantum Repeaters: The Role of Imperfect Local Operations in Quantum Communication},
  author = {Briegel, H.-J. and D\"ur, W. and Cirac, J. I. and Zoller, P.},
  journal = {Phys. Rev. Lett.},
  volume = {81},
  issue = {26},
  pages = {5932--5935},
  numpages = {0},
  year = {1998},
  month = {Dec},
  publisher = {American Physical Society},
  doi = {10.1103/PhysRevLett.81.5932},
  url = {https://link.aps.org/doi/10.1103/PhysRevLett.81.5932}
}

@article{Brendel:1999,
  title = {{Pulsed Energy-Time Entangled Twin-Photon Source for Quantum Communication}},
  author = {Brendel, J. and Gisin, N. and Tittel, W. and Zbinden, H.},
  journal = {Phys. Rev. Lett.},
  volume = {82},
  issue = {12},
  pages = {2594--2597},
  numpages = {0},
  year = {1999},
  month = {Mar},
  publisher = {American Physical Society},
  doi = {10.1103/PhysRevLett.82.2594},
  url = {https://link.aps.org/doi/10.1103/PhysRevLett.82.2594}
}

@article{Chen:2024,
  title = {Ultralow-Loss Integrated Photonics Enables Bright, Narrowband, Photon-Pair Sources},
  author = {Chen, Ruiyang and Luo, Yi-Han and Long, Jinbao and Shi, Baoqi and Shen, Chen and Liu, Junqiu},
  journal = {Phys. Rev. Lett.},
  volume = {133},
  issue = {8},
  pages = {083803},
  numpages = {8},
  year = {2024},
  month = {Aug},
  publisher = {American Physical Society},
  doi = {10.1103/PhysRevLett.133.083803},
  url = {https://link.aps.org/doi/10.1103/PhysRevLett.133.083803}
}

@article{Clausen:2011,
author = {Clausen, C and Usmani, I. and Bussi\`eres, F. and Sangouard, N. and Afzelius, M. and de Riedmatten, H. and Gisin, N.},
journal = {Nature},
pages = {508–511},
publisher = {Nature Publishing Group},
title = {{Quantum storage of photonic entanglement in a crystal}},
volume = {469},
month = {jan},
year = {2011},
url = {https://www.nature.com/articles/nature09662},
doi = {https://doi.org/10.1038/nature09662},
}

@article{Clauser:1969,
  title = {Proposed Experiment to Test Local Hidden-Variable Theories},
  author = {Clauser, John F. and Horne, Michael A. and Shimony, Abner and Holt, Richard A.},
  journal = {Phys. Rev. Lett.},
  volume = {23},
  issue = {15},
  pages = {880--884},
  numpages = {0},
  year = {1969},
  month = {Oct},
  publisher = {American Physical Society},
  doi = {10.1103/PhysRevLett.23.880},
  url = {https://link.aps.org/doi/10.1103/PhysRevLett.23.880}
}

@article{Duan:2001,
author = {L.-M. Duan and M. D. Lukin and J. I. Cirac and P. Zoller},
journal = {Nature},
issue = {6862},
pages = {413-418},
publisher = {Nature Publishing Group},
title = {{Long-distance quantum communication with atomic ensembles and linear optics}},
volume = {414},
year = {2001},
url = {https://doi.org/10.1038/35106500},
doi = {10.1038/35106500},
}

@article{Drever:1983,
author = {R. W. P. Drever and J. L. Hall and F. V. Kowalski and J. Hough and G. M. Ford and A. J. Munley and H. Ward},
journal = {Applied Physics B},
issue = {2},
pages = {97-105},
publisher = {Springer Science},
title = {Laser phase and frequency stabilization using an optical resonator},
volume = {31},
month = {June},
year = {1983},
url = {https://doi.org/10.1007/BF00702605},
doi = {10.1007/BF00702605},
}

@article{Halder:2007,
author = {M. Halder and A. Beveratos and N. Gisin and V. Scarani and C. Simon and H. Zbinden },
journal = {Nat. Phys.},
pages = {692–695},
publisher = {Nature Publishing Group},
title = {{Entangling independent photons by time measurement}},
volume = {3},
month = {aug},
year = {2007},
url = {https://www.nature.com/articles/nphys700#citeas},
doi = {10.1038/nphys700},
}

@article{Hanni:2025,
  title = {Heralded Entanglement of On-Demand Spin-Wave Solid-State Quantum Memories for Multiplexed Quantum Network Links},
  author = {H\"anni, Jonathan and Rodr\'{\i}guez-Moldes, Alberto E. and Appas, F\'elicien and Wengerowsky, Soeren and Lago-Rivera, Dario and Teller, Markus and Grandi, Samuele and de Riedmatten, Hugues},
  journal = {Phys. Rev. X},
  volume = {15},
  issue = {4},
  pages = {041003},
  numpages = {16},
  year = {2025},
  month = {Oct},
  publisher = {American Physical Society},
  doi = {10.1103/wvv1-6lg8},
  url = {https://link.aps.org/doi/10.1103/wvv1-6lg8}
}

@article{Hong:1987,
  title = {Measurement of subpicosecond time intervals between two photons by interference},
  author = {Hong, C. K. and Ou, Z. Y. and Mandel, L.},
  journal = {Phys. Rev. Lett.},
  volume = {59},
  issue = {18},
  pages = {2044--2046},
  numpages = {0},
  year = {1987},
  month = {Nov},
  publisher = {American Physical Society},
  doi = {10.1103/PhysRevLett.59.2044},
  url = {https://link.aps.org/doi/10.1103/PhysRevLett.59.2044}
}

@article{Huang:2010,
  title = {Heralding single photons without spectral factorability},
  author = {Huang, Yu-Ping and Altepeter, Joseph B. and Kumar, Prem},
  journal = {Phys. Rev. A},
  volume = {82},
  issue = {4},
  pages = {043826},
  numpages = {6},
  year = {2010},
  month = {Oct},
  publisher = {American Physical Society},
  doi = {10.1103/PhysRevA.82.043826},
  url = {https://link.aps.org/doi/10.1103/PhysRevA.82.043826}
}

@article{Husmann:2021,
author = {D. Husmann and L.-G. Bernier and M. Bertrand and D. Calonico and K. Chaloulos and G. Clausen and C. Clivati and J. Faist and E. Heiri and U. Hollenstein and A. Johnson and F. Mauchle and Z. Meir and F. Merkt and A. Mura and G. Scalari and S. Scheidegger and H. Schmutz and M. Sinhal and S. Willitsch and J. Morel},
journal = {Optics Express},
issue = {16},
pages = {24592-24605},
publisher = {Optica Publishing Group},
title = {{SI-traceable frequency dissemination at 1572.06  nm in a stabilized fiber network with ring topology}},
volume = {29},
year = {2021},
url = {https://doi.org/10.1364/OE.427921},
doi = {10.1364/OE.427921},
}

@article{Jiang:2007,
  title = {Distributed quantum computation based on small quantum registers},
  author = {Jiang, Liang and Taylor, Jacob M. and S\o{}rensen, Anders S. and Lukin, Mikhail D.},
  journal = {Phys. Rev. A},
  volume = {76},
  issue = {6},
  pages = {062323},
  numpages = {22},
  year = {2007},
  month = {Dec},
  publisher = {American Physical Society},
  doi = {10.1103/PhysRevA.76.062323},
  url = {https://link.aps.org/doi/10.1103/PhysRevA.76.062323}
}

@article{Kimble:2008,
author = {H. J. Kimble},
journal = {Nature},
pages = {1023–1030},
publisher = {Nature Publishing Group},
title = {{The quantum internet}},
volume = {453},
month = {jun},
year = {2008},
url = {https://doi.org/10.1038/nature07127},
doi = {https://doi.org/10.1038/nature07127},
}

@article{Kutluer:2019,
  title = {Time Entanglement between a Photon and a Spin Wave in a Multimode Solid-State Quantum Memory},
  author = {Kutluer, Kutlu and Distante, Emanuele and Casabone, Bernardo and Duranti, Stefano and Mazzera, Margherita and de Riedmatten, Hugues},
  journal = {Phys. Rev. Lett.},
  volume = {123},
  issue = {3},
  pages = {030501},
  numpages = {6},
  year = {2019},
  month = {Jul},
  publisher = {American Physical Society},
  doi = {10.1103/PhysRevLett.123.030501},
  url = {https://link.aps.org/doi/10.1103/PhysRevLett.123.030501},
}

@article{Legero:2003,
author = {Legero, T. and Wilk, T. and Kuhn, A. and Rempe, G.},
journal = {Appl. Phys. B},
pages = {797-802},
publisher = {Springer Science},
title = {{Time-resolved two-photon quantum interference}},
volume = {77},
year = {2003},
url = {https://doi.org/10.1007/s00340-003-1337-x},
doi = {10.1007/s00340-003-1337-x},
}

@article{Legero:2004,
  title = {Quantum Beat of Two Single Photons},
  author = {Legero, Thomas and Wilk, Tatjana and Hennrich, Markus and Rempe, Gerhard and Kuhn, Axel},
  journal = {Phys. Rev. Lett.},
  volume = {93},
  issue = {7},
  pages = {070503},
  numpages = {4},
  year = {2004},
  month = {Aug},
  publisher = {American Physical Society},
  doi = {10.1103/PhysRevLett.93.070503},
  url = {https://link.aps.org/doi/10.1103/PhysRevLett.93.070503}
}

@article{Liu:2023,
  title = {{Experimental Twin-Field Quantum Key Distribution over 1000 km Fiber Distance}},
  author = {Liu, Yang and Zhang, Wei-Jun and Jiang, Cong and Chen, Jiu-Peng and Zhang, Chi and Pan, Wen-Xin and Ma, Di and Dong, Hao and Xiong, Jia-Min and Zhang, Cheng-Jun and Li, Hao and Wang, Rui-Chun and Wu, Jun and Chen, Teng-Yun and You, Lixing and Wang, Xiang-Bin and Zhang, Qiang and Pan, Jian-Wei},
  journal = {Phys. Rev. Lett.},
  volume = {130},
  issue = {21},
  pages = {210801},
  numpages = {6},
  year = {2023},
  month = {May},
  publisher = {American Physical Society},
  doi = {10.1103/PhysRevLett.130.210801},
  url = {https://link.aps.org/doi/10.1103/PhysRevLett.130.210801}
}

@article{Llewellyn:2020,
author = {D. Llewellyn and Y. Ding and I. I. Faruque and S. Paesani and D. Bacco and R. Santagati and Y.-J. Qian and Y. Li and Y.-F. Xiao and M. Huber and M. Malik and G. F. Sinclair and X. Zhou and K. Rottwitt and J. L. O’Brien and J. G. Rarity and Q. Gong and L. K. Oxenlowe and J. Wang and M. G. Thompson },
journal = {Nat. Phys.},
pages = {148–153},
publisher = {Nature Publishing Group},
title = {{Chip-to-chip quantum teleportation and multi-photon entanglement in silicon}},
volume = {16},
month = {feb},
year = {2020},
url = {https://www.nature.com/articles/s41567-019-0727-x#citeas},
doi = {10.1038/s41567-019-0727-x},
}

@article{Main:2025,
author = {D. Maina and P. Drmota and D. P. Nadlinger and E. M. Ainley and A. Agrawal and B. C. Nichol and R. Srinivas and G. Araneda and D. M. Lucas },
journal = {Nature},
pages = {383–388},
publisher = {Nature Publishing Group},
title = {{Distributed quantum computing across an optical network link}},
volume = {638},
month = {feb},
year = {2025},
url = {https://www.nature.com/articles/s41586-024-08404-x#citeas},
doi = {10.1038/s41586-024-08404-x},
}

@article{Marchese:2023,
  title = {Large Baseline Optical Imaging Assisted by Single Photons and Linear Quantum Optics},
  author = {Marchese, Marta Maria and Kok, Pieter},
  journal = {Phys. Rev. Lett.},
  volume = {130},
  issue = {16},
  pages = {160801},
  numpages = {6},
  year = {2023},
  month = {Apr},
  publisher = {American Physical Society},
  doi = {10.1103/PhysRevLett.130.160801},
  url = {https://link.aps.org/doi/10.1103/PhysRevLett.130.160801}
}

@article{Moille:2025,
  title = {{Broadband Visible Wavelength Microcomb Generation In Silicon Nitride Microrings Through Air‐Clad Dispersion Engineering}},
  author = {G. Moille and D. Westly and R. Shrestha and K. Hoang and K. Srinivasan},
  journal = {Laser Photonics Rev.},
  volume = {19},
  pages = {2401746},
  year = {2025},
  month = {apr},
  doi = {10.1002/lpor.202401746},
  url = {https://onlinelibrary.wiley.com/doi/10.1002/lpor.202401746}
}

@article{Moody:2022,
author = {{G. Moody, V. J. Sorger, D. J. Blumenthal, et al.}},
journal = {J. Phys. Photonics},
number = {1},
pages = {012501},
publisher = {IOP Publishing Ltd},
title = {2022 {Roadmap} on integrated quantum photonics},
volume = {4},
month = {1},
year = {2022},
url = {https://iopscience.iop.org/article/10.1088/2515-7647/ac1ef4},
doi = {10.1088/2515-7647/ac1ef4},
}

@article{Muralidharan:2016,
title = {Optimal architectures for long distance quantum communication},
volume = {6},
issn = {2045-2322},
url = {https://www.nature.com/articles/srep20463},
doi = {10.1038/srep20463},
language = {en},
number = {1},
urldate = {2026-08-14},
journal = {Scientific Reports},
author = {Muralidharan, Sreraman and Li, Linshu and Kim, Jungsang and Lütkenhaus, Norbert and Lukin, Mikhail D. and Jiang, Liang},
month = feb,
year = {2016},
pages = {20463},
}

@article{Pal:2025,
author = {Arghadeep Pal and Alekhya Ghosh and Shuangyou Zhang and Toby Bi and Masoud Kheyri and Haochen Yan and Yaojing Zhang and Pascal Del\'Haye},
journal = {arXiv},
volume = {2505.01352},
title = {{Hybrid Nonlinear Effects in Photonic Integrated Circuits}},
year = {2025},
}

@article{Peres:1996,
  title = {Separability Criterion for Density Matrices},
  author = {Peres, Asher},
  journal = {Phys. Rev. Lett.},
  volume = {77},
  issue = {8},
  pages = {1413--1415},
  numpages = {0},
  year = {1996},
  month = {Aug},
  publisher = {American Physical Society},
  doi = {10.1103/PhysRevLett.77.1413},
  url = {https://link.aps.org/doi/10.1103/PhysRevLett.77.1413}
}

@article{Pfeiffer:2016,
author = {M. H. P. Pfeiffer and A. Kordts and V. Brasch and M. Zervas and M. Geiselmann and J. D. Jost and T. J. Kippenberg},
journal = {Optica},
issue = {1},
pages = {20-25},
publisher = {Optica Publishing Group},
title = {{Photonic Damascene process for integrated high-Q microresonator based nonlinear photonics }},
volume = {3},
year = {2016},
url = {https://doi.org/10.1364/OPTICA.3.000020},
doi = {10.1364/OPTICA.3.000020},
}

@article{Pirandola:2017,
author = {S. Pirandola and R. Laurenza and C. Ottaviani and L. Banchi },
journal = {Nat. Commun.},
issue = {},
pages = {15043},
publisher = {Nature Publishing Group},
title = {Fundamental limits of repeaterless quantum
communications},
volume = {8},
year = {2017},
url = {https://doi.org/10.1038/ncomms15043},
doi = {10.1038/ncomms15043},
}

@article{Pirandola:2020,
author = {S. Pirandola and U. L. Andersen and L. Banchi and M. Berta and D. Bunandar and R. Colbeck and D. Englund and T. Gehring and C. Lupo and C. Ottaviani and J. L. Pereira and M. Razavi and J. Shamsul Shaari and M. Tomamichel and V. C. Usenko and G. Vallone and P. Villoresi and P. Wallden},
journal = {Adv. Opt. Photon.},
number = {4},
pages = {1012--1236},
publisher = {Optica Publishing Group},
title = {Advances in quantum cryptography},
volume = {12},
month = {Dec},
year = {2020},
url = {https://opg.optica.org/aop/abstract.cfm?URI=aop-12-4-1012},
doi = {10.1364/AOP.361502},
}

@article{Rivera:2021,
author = {D. Lago-Rivera and S. Grandi and J. V. Rakonjac and A. Seri and H. de Riedmatten},
journal = {Nature},
pages = {37-40},
publisher = {Nature Publishing Group},
title = {Telecom-heralded entanglement between multimode solid-state quantum memories},
volume = {594},
month = {6},
year = {2021},
url = {https://doi.org/10.1038/s41586-021-03481-8},
doi = {10.1038/s41586-021-03481-8},
}

@article{Rogers:2016,
   author = {Steven Rogers and Daniel Mulkey and Xiyuan Lu and Wei C. Jiang and Qiang Lin},
   doi = {10.1021/acsphotonics.6b00423},
   issn = {23304022},
   issue = {10},
   journal = {ACS Photonics},
   month = {10},
   pages = {1754-1761},
   publisher = {American Chemical Society},
   title = {High Visibility Time-Energy Entangled Photons from a Silicon Nanophotonic Chip},
   volume = {3},
   year = {2016},
}

@article{Samara:2019,
author = {F. Samara and A. Martin and C. Autebert and M. Karpov and T. J. Kippenberg and H. Zbinden and R. Thew},
journal = {Opt. Express},
issue = {14},
pages = {19309-19318},
title = {{High-rate photon pairs and sequential Time-Bin entanglement with $\text{Si}_{3}\text{N}_{4}$ microring resonators}},
volume = {27},
year = {2019},
url = {https://doi.org/10.1364/OE.27.019309},
doi = {10.1364/OE.27.019309},
}

@article{Samara:2021,
author = {F. Samara and N. Maring and A. Martin and A. S. Raja and T. J. Kippenberg and H. Zbinden and R. Thew},
journal = {Quantum Sci. Technol.},
pages = {145024},
title = {Entanglement swapping between independent and asynchronous integrated photon-pair sources},
volume = {6},
year = {2021},
url = {https://iopscience.iop.org/article/10.1088/2058-9565/abf599},
doi = {10.1088/2058-9565/abf599},
}

@article{Sanchez:2026,
author = {T. Sanchez Mejia and L. Nicolas and A. G. Rodriguez and P. Goldner and M. Afzelius},
journal = {Quantum Sci. Technol.},
pages = {015004},
title = {{Broadband and long-duration optical memory in $^{171}\text{Yb}^{3+}:\text{Y}_{2}\text{SiO}_{5}$}},
volume = {11},
year = {2026},
url = {https://iopscience.iop.org/article/10.1088/2058-9565/ae1bd0},
doi = {10.1088/2058-9565/ae1bd0},
}

@article{Sangouard:2011,
author = {N. Sangouard and C. Simon and H. de Riedmatten and N. Gisin},
journal = {Rev. Mod. Phys.},
pages = {33-80},
publisher = {American Physical Society},
title = {Quantum repeaters based on atomic ensembles and linear optics},
volume = {83},
month = {3},
year = {2011},
url = {https://journals.aps.org/rmp/abstract/10.1103/RevModPhys.83.33},
doi = {10.1103/RevModPhys.83.33},
}

@article{Singh:2025,
author = {A. Singh and A. Sethia and L. Esmaeilifar and R. Valivarthi and N. Sinclair and M. Spiropulu and D. Oblak},
journal = {arXiv},
volume = {2507.08102},
title = {{Photonic quantum information with time-bins: Principles and applications}},
year = {2025},
}

@article{Stas:2026,
author = {P.-J. Stas and Y.-C. Wei and M. Sirotin and Y. Q. Huan and U. Yazlar and F. Abdo Arias and E. Knyazev and G. Baranes and B. Machielse and S. Grandi and D. Riedel and J. Borregaard and H. Park and M. Lon\car and A. Suleymanzade and M. D. Lukin },
journal = {Nature},
pages = {326–332},
publisher = {Nature Publishing Group},
title = {{Entanglement-assisted non-local optical interferometry in a quantum network}},
volume = {651},
month = {mar},
year = {2026},
url = {https://www.nature.com/articles/s41586-026-10171-w#citeas},
doi = {10.1038/s41586-026-10171-w},
}

@article{Takeoka:2014,
author = {M. Takeoka and S. Guha and M. M. Wilde},
journal = {Nat. Commun.},
pages = {5235},
publisher = {Nature Publishing Group},
title = {Fundamental rate-loss tradeoff for optical quantum key distribution},
volume = {5},
year = {2014},
url = {https://doi.org/10.1038/ncomms6235},
doi = {10.1038/ncomms6235},
}

@article{Tittel:2010,
  title = {{Photon-echo quantum memory in solid state systems}},
  author = {W. Tittel and M. Afzelius and T. Chaneli\'ere and R. L. Cone and S. Kr$\ddot{\text{o}}$ll and S. A. Moiseev and M. Sellars},
  journal = {Laser Photonics Rev.},
  volume = {4},
  number = {2},
  pages = {244–267},
  year = {2010},
  month = {February},
  doi = {https://doi.org/10.1002/lpor.200810056},
  url = {10.1002/lpor.200810056}
}

@article{Tittel:2025,
author = {W. Tittel and M. Afzelius and A. Kinos and L. Rippe and A. Walther},
journal = {Quantum Sci. Technol.},
pages = {033002},
title = {{Quantum networks using rare-earth ions}},
volume = {10},
year = {2025},
url = {https://iopscience.iop.org/article/10.1088/2058-9565/addd93},
doi = {10.1088/2058-9565/addd93},
}

@article{Wehner:2018,
author = {Stephanie Wehner and D. Elkouss and R. Hanson},
title = {{Quantum internet: A vision for the road ahead}},
journal = {Science},
volume = {362},
issue = {6412},
pages = {eaam9288},
year = {2018},
doi = {10.1126/science.aam9288},
URL = {},
}

@article{Zektzer:2024,
author = {R. Zektzer and X. Lu and K. T. Hoang and R. Shrestha and S. Austin and F. Zhou and A. Chanana and G. Holland and D. Westly and P. Lett and A. V. Gorshkov and K. Srinivasan},
journal = {Optica},
issue = {10},
pages = {1376-1384},
publisher = {Optica Publishing Group},
title = {{Strong interactions between integrated microresonators and alkali atomic vapors: towards single-atom, single-photon operation}},
volume = {11},
year = {2024},
url = {https://doi.org/10.1364/OPTICA.525689},
doi = {10.1364/OPTICA.525689},
}

@article{Zukowski:1993,
  title = {``Event-ready-detectors'' Bell experiment via entanglement swapping},
  author = {\ifmmode \dot{Z}\else \.{Z}\fi{}ukowski, M. and Zeilinger, A. and Horne, M. A. and Ekert, A. K.},
  journal = {Phys. Rev. Lett.},
  volume = {71},
  issue = {26},
  pages = {4287--4290},
  numpages = {0},
  year = {1993},
  month = {Dec},
  publisher = {American Physical Society},
  doi = {10.1103/PhysRevLett.71.4287},
  url = {https://link.aps.org/doi/10.1103/PhysRevLett.71.4287}
}

\end{document}